\documentclass[11pt]{article}
\usepackage[margin=1in]{geometry}
\usepackage[numbers,sort&compress]{natbib}
\usepackage{times}
\usepackage{hyperref}
\usepackage{url}
\usepackage{graphicx}
\usepackage{booktabs}
\usepackage{amsmath,amssymb}
\usepackage{enumitem}
\usepackage{xcolor}
\usepackage{caption}
\usepackage{float}

\usepackage{amsmath,amsfonts,bm}

\def\eqref#1{equation~\ref{#1}}

\def\1{\bm{1}}

\DeclareMathAlphabet{\mathsfit}{\encodingdefault}{\sfdefault}{m}{sl}
\SetMathAlphabet{\mathsfit}{bold}{\encodingdefault}{\sfdefault}{bx}{n}

\setlist{nosep,leftmargin=*}
\title{Toward Controlling Biology with Language:\\
Offline Learning of Prompt-Conditioned Interventions for Cells, Organoids, and Biobots}

\author{
Nam H. Le\thanks{Department of Computer Science and Vermont Complex Systems Institute,
University of Vermont. Correspondence: \texttt{nam.le@uvm.edu}} \and
Douglas Blackiston\thanks{Allen Discovery Center, Tufts University.} \and
Michael Levin\footnotemark[2] \and
Josh Bongard\footnotemark[1]
}

\date{}

\begin{document}
\maketitle

\begin{abstract}
Natural language has become an interface to code, images and molecules, but not to living
matter: a stimulus applied to cells has no fixed linguistic meaning, and every paired example
of an instruction and its outcome costs a wet-lab experiment. We ask how far an existing
archive of experiments can take this interface, using xenobots, motile multicellular
constructs made of frog cells, stimulated by a single electrical pulse of varying duration.
We re-track 155 archived recordings with two independent methods, measure what the pulse
does from each bot's own behaviour before and after it, and separate behaviours the pulse
controls from behaviours it does not. Longer pulses stop bots and short pulses leave them
moving, sometimes faster; turning direction and starting from rest do not depend on duration,
and restarting at most weakly. We then replace hand-made behaviour labels with a blind language critic: fresh
sessions of a multimodal language model read plots of each bot and score how well its
behaviour matches each of 176 instructions. The critic agrees with the measured labels
whenever its plots contain the information the labels are defined on, and dose curves
learned from its scores match those learned from the measurements. A prompt-to-intervention
model trained only on critic scores maps unseen instructions, including negations, to the
right end of the dose range on unseen bots, as well as the same model trained on measured
labels. The archive therefore supports a calibrated, refusing language interface for the
three behaviours a pulse can set, and identifies, rather than hides, the behaviours it cannot.
\end{abstract}

\section{Introduction}
\label{sec:intro}

A growing set of technical systems can be driven by describing a goal in words: language
models write code \citep{chen2021evaluating}, plan actions \citep{huang2022zeroshotplanners}
and assist scientific work \citep{zheng2025large}. Living systems are a harder target.
Cells, tissues and engineered organisms respond to physical and chemical interventions
\citep{levin2021bioelectric, emiliani2022optogenetics}, but an intervention such as an
electrical pulse has no closed-form meaning in language, and the data needed to learn one,
triples of instruction, intervention and outcome, cost an experiment each.

Laboratories already hold archives of interventions and their recorded outcomes, collected for
other purposes. This paper studies what such an archive can support, using xenobots,
self-assembled motile constructs of \textit{Xenopus} cells with no nervous system
\citep{kriegman2021kinematic, blackiston2021cellular}, stimulated by a single electrical pulse
whose duration varies between recordings. Three questions organise the work:
\begin{enumerate}
\item \textbf{What does the intervention do, and what can it control?} Before any language is
involved, the outcomes must be measured reliably, and behaviours that depend on the
intervention must be separated from behaviours that occur at the same rate whatever is done.
\item \textbf{Can a language model replace hand-made labels?} Defining each behaviour by hand
does not scale to open-ended instructions. We test whether a blind language critic, shown
plots of a bot and an instruction, scores the bot in agreement with independent measurements,
and when it does not.
\item \textbf{Can an instruction be mapped to an intervention that generalises?} We learn a
mapping from the text of an instruction to a dose--response curve, and test it on
instructions and bots never seen in training.
\end{enumerate}

\paragraph{Contributions.}
\begin{itemize}
\item A re-tracked and audited version of the archive (155 recordings, two independent
trackers), with a measurement protocol whose baseline length is chosen by reliability on
unstimulated data, not by outcome (Sections~\ref{sec:data}, \ref{sec:outcomes}).
\item A behaviour map of the intervention: longer pulses stop bots (stopping within 10 minutes
rises from 0/22 bots at pulses of 30\,s or less to 17/19 above 220\,s), short pulses leave
them moving and sometimes faster, and turning direction and starting from rest do not depend
on pulse duration, restarting at most weakly (Section~\ref{sec:res-pulse}).
\item A blind language critic, run as one fresh model session per image with a fixed template
and fully logged, whose scores agree with measured labels (AUC 0.85--0.98) once its plots show
the information each label uses; where it disagreed, it exposed weaknesses in our own labels
(Sections~\ref{sec:critic}, \ref{sec:res-critic}).
\item A prompt-to-intervention (P2I) model that outputs a dose--response curve for any
instruction. Trained only on critic scores, it chooses the right dose for 26 of 26 unseen
instructions (10 of 10 negations) and predicts the measured behaviour of unseen bots as well
as the same model trained on measured labels (Sections~\ref{sec:p2i}, \ref{sec:res-p2i}).
\end{itemize}

The archive contains no sham (unstimulated) control recordings. All claims are therefore
predictive and dose--response claims within the archive's protocol; Section~\ref{sec:limits}
states what this rules out.

\section{Related work}
\label{sec:related}

\paragraph{Controlling living constructs.} Xenobots are motile assemblies of frog epidermal
cells whose behaviour can be shaped by design and by stimulation
\citep{kriegman2021kinematic, blackiston2021cellular}. Electrical and optogenetic
interventions change behaviour in many living systems \citep{zajdel2020scheepdog,
levin2021bioelectric, emiliani2022optogenetics}; controlling them is usually done by hand,
parameter by parameter. Earlier work on language-directed control used simulated cell
collectives with a vision-language reward \citep{le2025giving, Le2025zapgpt}; here every
outcome is a recorded living organism.

\paragraph{Language and vision models as judges.} Vision-language models have been used as
zero-shot reward models \citep{rocamonde2023vision, radford2021learning}, and language models
as judges of other models' outputs \citep{zheng2023judging}. Such judges are rarely validated
against an independent measurement of the same property; we validate the critic against
measured outcomes on every behaviour, report where it fails, and trace each failure.

\paragraph{Sentence encoders and negation.} Frozen sentence encoders
\citep{reimers2019sentence} give strong general-purpose embeddings, but negation is a known
weakness of language representations \citep{ettinger2020bert}. Encoders fine-tuned on natural
language inference are a natural choice when instructions such as ``do not stop it'' must be
told apart from ``stop it''.

\paragraph{Dose--response and survival analysis.} Probability of a response against log dose
is classically modelled with a logistic or probit curve \citep{finney1971probit}; time to an
event with censoring by Kaplan--Meier curves and the proportional-hazards model
\citep{cox1972regression}. The P2I model outputs the parameters of a logistic dose--response
curve, so its predictions read directly as an effective dose and a width.

\section{Data}
\label{sec:data}

\paragraph{The archive.}
The data are an existing archive of electrical-stimulation experiments on xenobots, recorded
for purposes other than this study. No new experiment was run for this paper. Each recording
follows one bot through three phases: an unstimulated phase (PRE), a single direct-current
pulse of 20\,mA, and the period after the pulse (POST). A static camera images the dish from
above every 5\,s at $9152 \times 6944$ pixels; the bot is about 76--110 pixels across,
roughly 1\% of the frame width. Pulse duration $d$ is the only stimulation parameter that
varies systematically, from 1 to 327\,s (86 distinct values); the field angle is fixed in all
but eight recordings and is not used. Each recording is a different bot, so recordings are
independent biological replicates. The archive contains no sham recordings.

\paragraph{From 221 archive entries to 155 recordings.}
Table~\ref{tab:dataflow} lists every step from the archive to the analysis set. Of 221 archive
entries, 163 contain a stimulation command and an operator click on the bot in the first
frame. Eight are excluded: five in which both trackers lose the bot, and three whose PRE or
POST phase is shorter than 10 minutes. The remaining 155 recordings are used throughout; PRE
lasts a median 43\,min (range 16--114) and POST 42\,min (17--340).

\begin{table}[t]
\centering
\small
\caption{Data flow. Every count is a recording, that is, one bot.}
\label{tab:dataflow}
\begin{tabular}{lrl}
\toprule
step & recordings & note \\
\midrule
archive entries & 221 & \\
with a stimulation command and an initial click & 163 & \\
excluded: both trackers lose the bot & $-5$ & 078, 079, 285, 392, 394 \\
excluded: PRE or POST shorter than 10\,min & $-3$ & 224, 228, 230 \\
\textbf{analysis set} & \textbf{155} & \\
\quad moving at pulse onset & 121 & speed in the last 3\,min of PRE $\geq 0.1$\,px/s \\
\quad still at pulse onset & 34 & used only for ``starts moving'' \\
\bottomrule
\end{tabular}
\end{table}

\paragraph{Tracking.}
The bot covers 0.01\% of the image. Bubbles, electrode edges, rim shadows and debris form
dark, bot-sized regions; the bot moves a median 0.04 body lengths per 10\,s, so a nearly
motionless bot sits at the resolution limit; illumination is uneven. A general-purpose
segmentation tracker \citep{kirillov2023segment} fails silently on this footage: on 23.8\% of
frames it does not find the bot, and in thirteen recordings on more than 80\%. Rendering the
failure frames shows why: the segmenter returns a thin band across the dish, a pixel-count
test accepts it, the position jumps by about 3,000 pixels, and a local search never
recovers, while the bot remains visible in place. Every recording was therefore tracked
with two independent methods (Appendix~\ref{app:tracking}):
\begin{itemize}
\item \textbf{Robust PCA (primary).} Each recording's frame stack is decomposed into a
low-rank background and a sparse foreground \citep{candes2011robust} (inexact augmented
Lagrangian \citep{lin2010augmented}, 1/16 resolution). The bot is followed in the sparse
layer from the operator click. A bot that stops is absorbed into the background within
minutes; its last position is then held, and the raw frame is checked for a dark object
there.
\item \textbf{Click-initialised classical tracker (check).} A darkness map at 1/4
resolution, candidate filters calibrated to this bot's own size, darkness and surroundings
(which reject hollow bubbles and objects on the rim), temporal gating, and global
re-acquisition after three missed frames.
\end{itemize}

Over all 163 recordings the median distance between the two trackers is 0.13, 0.13 and 0.17
body lengths in PRE, during the pulse and in POST; in 122 recordings at least 95\% of frames
lie within half a body length; baseline speeds correlate at Spearman 0.97
(Figure~\ref{fig:a-trackers}). Most of the remaining disagreement is a constant offset of the
held robust-PCA position (0.55--0.73 body lengths). Every recording in which robust PCA lost
the bot was inspected by eye: in ten, the bot had stopped and stays at the held position; in
six, the classical tracker follows the bot through the lost frames and is used there; in
five, neither tracker can be trusted and the recording is excluded. The final track is
robust PCA, unchanged for 139 of the 155 recordings and repaired only in the lost frames of
the other 16.

\paragraph{Frames used.}
All measurements use PRE and POST only. Frames recorded during the pulse are never used to
define a behaviour, to show the critic, or to train a model: during the pulse the electrodes
produce bubbles and the field acts on the dish, so motion in those frames is not the bot's
own behaviour. A bot that stops during the pulse enters POST stopped, which is recorded as its
state after the pulse.

\section{Measured outcomes}
\label{sec:outcomes}

\paragraph{Baseline, chosen by reliability.}
Bots drift: their speed changes over tens of minutes without any stimulation. A baseline must
therefore be long enough to be reliable and short enough to describe the bot at the moment of
the pulse. We chose its length on PRE frames only, before looking at any outcome, by the
agreement of two adjacent windows at the end of PRE (intraclass correlation, ICC(3,1)
\citep{shrout1979intraclass}) and their natural drift. Three minutes gives ICC 0.960, 90\% of
bots within a factor 1.45 of drift, and a usable baseline for 98\% of bots
(Figure~\ref{fig:a-baseline}); longer windows lose reliability (10\,min: ICC 0.850). The state
of a bot at pulse onset is its mean log speed over the last 3 minutes of PRE; a bot is
\emph{moving at onset} if this is at least 0.1\,px/s (121 bots) and \emph{still} otherwise (34).
All dose--response results were repeated with 1, 2 and 5-minute baselines and do not change
(Appendix~\ref{app:sensitivity}). Each bot's \emph{signature} (typical speed, turning direction,
curvature, speed spread and trend) is measured over the whole PRE phase.

\paragraph{Outcomes.}
POST is cut into consecutive 3-minute windows starting at the end of the pulse. The
\emph{effect} in window $k$ is the mean log speed in that window minus the state at onset; the
\emph{change over 0--15\,min} is the mean of the first five effects. A bot \emph{stops} at the
first 3-minute window whose mean log speed falls below 0.1\,px/s; time to stop is measured
from the end of the pulse and censored at the end of the recording. Table~\ref{tab:behaviours}
defines the eight behaviours used below. Turning behaviours use a turning baseline of the same
length (ICC 0.946 for turning direction) and are measurable only while the bot moves.

\begin{table}[t]
\centering
\small
\caption{Behaviour definitions. All are measured against the bot's own state at pulse onset;
frames during the pulse are not used.}
\label{tab:behaviours}
\begin{tabular}{lll}
\toprule
behaviour & bots it is defined on & measured label \\
\midrule
stops (STOP) & moving at onset (121) & stops within 10\,min of the pulse \\
keeps moving (KEEPS\_MOVING) & moving at onset (121) & does not stop within 10\,min \\
faster (FASTER) & moving at onset (121) & change over 0--15\,min $>0$ \\
slower, still moving (SLOWER\_MOVING) & moving at onset (121) & change $<0$, no stop within 15\,min \\
stops, then moves again (STOP\_RESTART) & moving, and stopped (81) & a later window $\geq0.1$\,px/s \\
starts moving (STILL\_STARTS) & still at onset (34) & any POST window $\geq0.1$\,px/s \\
turning direction reversed (TURN\_REVERSED) & with turning data (59) & sign of turning flips in 0--15\,min \\
turning direction kept (TURN\_KEPT) & with turning data (59) & complement of the above \\
\bottomrule
\end{tabular}
\end{table}

\paragraph{Dose--response and prediction.}
Because each recording is a different bot, inference is between bots across doses. Proportions
by pulse group are reported with exact Clopper--Pearson intervals and tested for a trend with
the Cochran--Armitage test \citep{armitage1955tests}; time to stop with Kaplan--Meier curves and
a Cox model on log duration \citep{cox1972regression}. Prediction on unseen bots uses
leave-one-out cross-validation for continuous outcomes and leave-pair-out AUC for binary ones
\citep{airola2011experimental}, with every step refitted inside each fold and 200 label
permutations of the whole pipeline for $p$-values. The probability of each behaviour against
dose is a logistic curve in log duration with a weak Gaussian prior on the slope, the form P2I
also outputs (Section~\ref{sec:p2i}).

\paragraph{Why no constructed control.}
Without sham recordings, one could construct a reference by placing fake pulse onsets inside
PRE. We tested this and rejected it: on PRE data alone, per-bot null distributions flag
21--23\% of fake onsets per side as significant and parametric event studies 27--62\%,
against a nominal 5\%, because bots change character over tens of minutes
(Appendix~\ref{app:placebo}). The shortest pulses are not a matched control either: they come
with longer PRE phases and faster bots. We therefore make no claim against ``no pulse''; all
claims are associations with dose within the archive and predictions on unseen bots.

\section{A blind language critic}
\label{sec:critic}

Defining every behaviour by hand, as in Table~\ref{tab:behaviours}, does not scale to
open-ended instructions. We test whether a language model can take this role: given plots of
one bot and an instruction, it scores how well the bot's behaviour after the pulse, compared
with its own behaviour before, matches the instruction.

\paragraph{What the critic sees.}
The critic sees plots, not numbers, and never sees the pulse duration, the batch identity or
any label. Frames during the pulse are removed. Three views were used (Figure~\ref{fig:critic}):
\begin{itemize}
\item \textbf{View B} shows speed every 10\,s and the path, for the last 3 minutes of PRE and
the first 15 minutes of POST: the data the STOP, KEEPS\_MOVING, FASTER and SLOWER\_MOVING
labels use. Nothing on the plot marks these windows.
\item \textbf{View C} shows speed on a log scale, signed turning every 10\,s, and the path with
arrows for the direction of travel, for the last 3 minutes of PRE and the whole POST: the
information STOP\_RESTART, STILL\_STARTS and the turning labels use.
\item \textbf{View A} (pilot only) shows the whole PRE and POST.
\end{itemize}

\paragraph{Protocol.}
Each image is scored by a fresh session of a commercial multimodal language model (Claude
Opus 5.5, Anthropic) that has no context beyond a fixed template (Appendix~\ref{app:templates}).
The judge first describes the bot in a few sentences, then scores every instruction from 0
(clearly does not match) to 100 (clearly matches), with a short reason. Images carry random
codes; the code-to-recording key, the exact text each judge received, the instruction order
(randomised per image), every raw response, and SHA-256 hashes of all files are stored.
View B was scored on all 155 bots for all 176 instructions (27,280 scores), view C on all 155
bots for the 90 instructions of its five behaviours (13,950 scores).

\paragraph{Validation.}
The critic is validated against the measured labels, on the bots each label is defined on, by
the AUC of the critic's score for an instruction against the measured label of the behaviour
the instruction requests. Negation consistency is checked between opposite instructions.

\section{Prompt-to-intervention mapping}
\label{sec:p2i}

\paragraph{Instructions.}
We wrote 176 instructions in nine classes: one per behaviour of Table~\ref{tab:behaviours} and
an out-of-scope class (e.g.\ ``make it photosynthesise''). 114 are used for training and 62
are held out, 20 of them negated (e.g.\ ``never let it come to rest'' requests KEEPS\_MOVING).
Instructions were written and labelled before any model was trained.

\paragraph{Model.}
An instruction $g$ is embedded by a frozen sentence encoder, nli-mpnet-base-v2
\citep{reimers2019sentence, song2020mpnet}, chosen for its training on natural language
inference. A small head maps the embedding to two numbers $(\beta_0, \beta_1)$, which define a
logistic dose--response curve
\begin{equation}
P(\text{behaviour requested by } g \mid d) = \sigma\!\left(\beta_0 + \beta_1 z(d)\right),
\qquad z(d) = \frac{\log d - m}{s},
\label{eq:curve}
\end{equation}
with $m, s$ the mean and standard deviation of log duration over the bots. Equivalently,
$P = \sigma\big((\log d - \mu)/\tau\big)$ with $\mu = m - s\beta_0/\beta_1$, the log of the
duration at which the behaviour has probability one half, and $\tau = s/|\beta_1|$, the width;
the sign of $\beta_1$ says whether longer or shorter pulses favour the behaviour. The main head
is a one-hidden-layer MLP (768 to 64, ReLU, dropout 0.1, to 2); a linear head and LoRA
fine-tuning of the encoder \citep{hu2022lora} are compared in Section~\ref{sec:res-ablation}.

\paragraph{Training targets.}
Each training instruction is paired with every training bot its class is defined on. The
target for a pair is either the \emph{critic score} of that instruction on that bot divided
by 100, or the \emph{measured label} (0 or 1) of the behaviour the instruction requests. The
loss is the cross-entropy between the target and Equation~\ref{eq:curve}, with weight decay;
targets are never clipped. Training on critic scores uses no measured label at any step.

\paragraph{Planning.}
For an instruction, the planned duration is the one in the archive's range (1--330\,s) that
maximises the predicted probability. Because the curve is monotone, this is always one end of
the range; a planner that returns the shortest duration reaching a target probability, and
refuses when none does, is discussed in Section~\ref{sec:limits}.

\paragraph{Evaluation.}
Generalisation is measured along two axes at once. Bots are split into five folds; the model
is trained on the training instructions and four folds of bots, and evaluated on every
combination of seen or unseen instructions with seen or unseen bots. In each cell we report
the AUC of the predicted probability against the measured behaviour, always the measured
label whatever the training target, averaged over behaviours. We also report whether the
planned duration falls at the end of the range where the requested behaviour is most likely
(330\,s for STOP, 1\,s for KEEPS\_MOVING and FASTER), separately for negated instructions.

\section{Results}
\label{sec:results}

\begin{figure}[t]
\centering
\includegraphics[width=\textwidth]{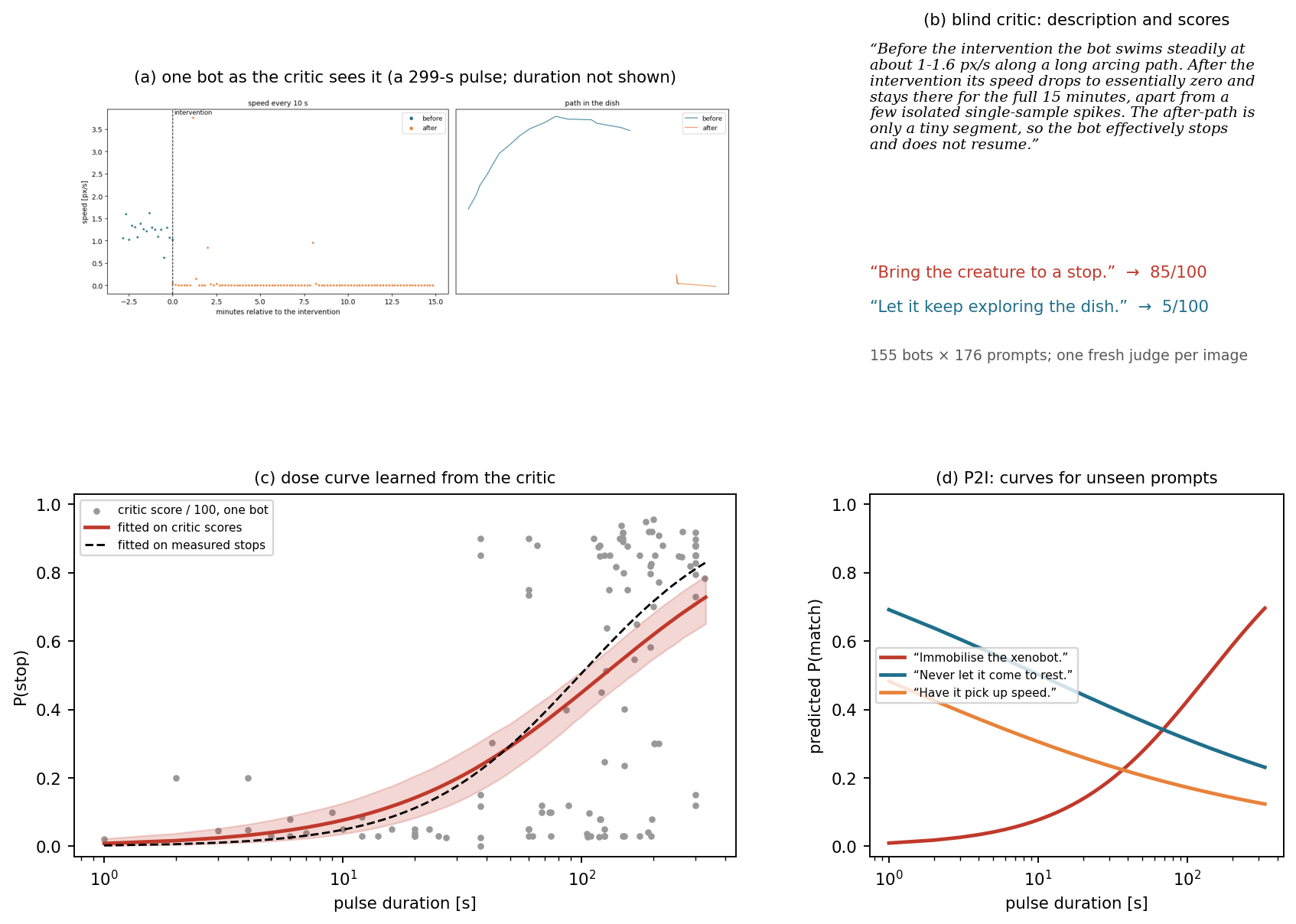}
\caption{From one archived bot to a language interface. (a) The image a blind critic receives
for one bot: speed every 10\,s and path, before (blue) and after (orange) the pulse; the pulse
duration is not shown. (b) The critic's description of this bot and two of its 176 scores.
(c) STOP dose curve learned from critic scores alone (red, 95\% bootstrap band) and from
measured stops (dashed), over the 121 bots moving at onset (grey: mean critic score per bot).
(d) Dose curves predicted by a P2I head (linear, trained on critic scores of the training
instructions) for three instructions it never saw.}
\label{fig:overview}
\end{figure}

\subsection{What the pulse does}
\label{sec:res-pulse}

\begin{figure}[t]
\centering
\includegraphics[width=\textwidth]{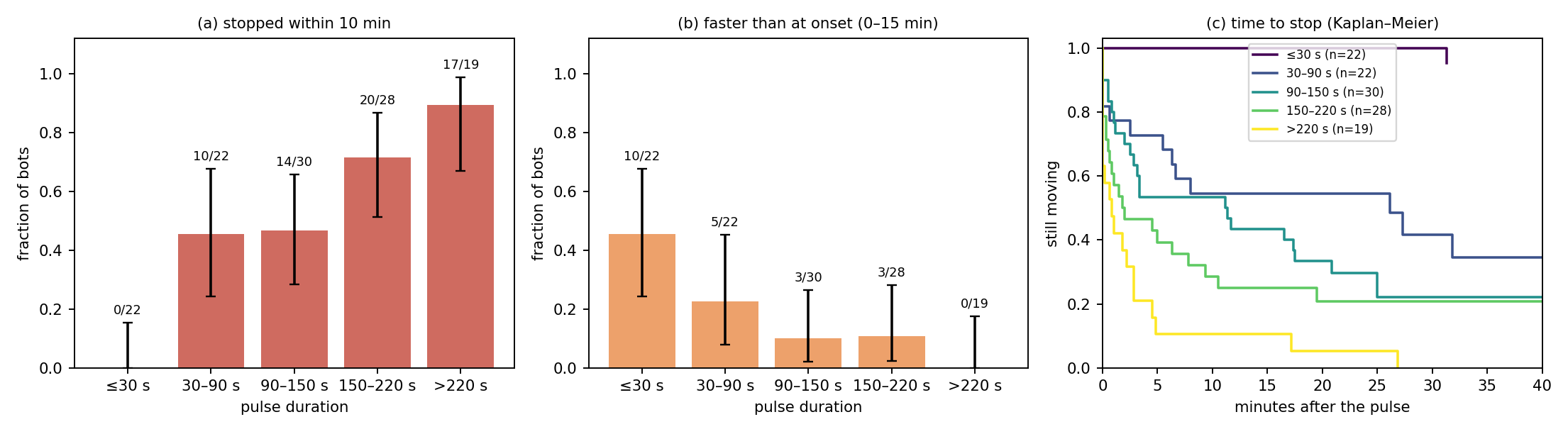}
\caption{What the pulse does, 121 bots moving at onset. (a) Fraction stopped within 10 minutes
after the pulse, and (b) fraction faster than at onset over 0--15 minutes, by pulse duration
(95\% Clopper--Pearson intervals; counts above bars). (c) Time to stop, Kaplan--Meier, censored
at the end of each recording.}
\label{fig:dose}
\end{figure}

Among the 121 bots moving at onset, the probability of stopping within 10 minutes rises with
pulse duration, from 0/22 for pulses of 30\,s or less to 10/22, 14/30, 20/28 and 17/19 in the
longer groups (Cochran--Armitage $z = 6.07$, $p = 10^{-9}$; Figure~\ref{fig:dose}a). Time to
stop shortens with dose (Figure~\ref{fig:dose}c): each $e$-fold increase in duration multiplies
the hazard of stopping by 2.60 (95\% CI 1.85--3.66, $p = 4\times10^{-8}$). Neither the state
at onset nor any component of the bot's PRE signature predicts stopping ($p \geq 0.11$).
Being faster than at onset falls with dose, from 10/22 at 30\,s or less to 0/19 above
220\,s ($z = -4.26$, $p = 2\times10^{-5}$; Figure~\ref{fig:dose}b), and the change over
0--15\,min falls with log duration (Spearman $-0.57$, $p = 7\times10^{-12}$).

On unseen bots, duration alone predicts stopping within 10 minutes with leave-pair-out AUC
0.83 ($p = 0.005$, permutation); adding the state at onset or the PRE signature does not
improve it (0.81, 0.79). The size of the change over 0--15\,min is predicted with
leave-one-out $R^2 = 0.26$ from duration and 0.42 once the state at onset is added; the
signature adds nothing beyond that. Whether a bot stops is set by the dose; how much it slows
depends also on its state when the pulse arrives.

Other behaviours were measured with the same machinery (Table~\ref{tab:rare}). Bots that
stop restart less often after long pulses (trend $p = 0.06$); still bots start moving at a
rate unrelated to dose ($p = 0.25$); turning direction is kept at the natural rate wherever it
can be measured and is reversed in about 17\% of bots at every dose ($p = 0.98$). A search
for behaviour that appears only after the pulse, by the distance of each 3-minute POST segment
to the repertoire of PRE segments of other bots, found fewer unusual POST segments (0.7\%)
than expected by chance (1\%) (Appendix~\ref{app:discovery}). Within this archive the pulse
sets three behaviours, stopping, continuing to move, and moving faster, and leaves the others
to the bot.

\begin{table}[t]
\centering
\small
\caption{Every behaviour against dose: events, logistic probability at three durations
(95\% bootstrap interval), and the Cochran--Armitage trend across pulse groups.}
\label{tab:rare}
\begin{tabular}{lrlll}
\toprule
behaviour & events & $P$ at 5\,s / 60\,s / 240\,s & dose trend & controlled \\
\midrule
stops within 10\,min & 61/121 & 0.02 / 0.35 / 0.76 & rises, $p=10^{-9}$ & yes \\
faster than at onset & 21/121 & 0.48 / 0.18 / 0.09 & falls, $p=2\times10^{-5}$ & yes (short pulses) \\
slower, keeps moving & 37/121 & 0.64 / 0.33 / 0.20 & falls, $p=9\times10^{-4}$ & weakly \\
stops, then moves again & 36/81 & 0.83 / 0.56 / 0.38 (wide) & falls, $p=0.06$ & no \\
still bot starts moving & 10/34 & 0.62 / 0.36 / 0.24 (wide) & none, $p=0.25$ & no \\
turning direction reversed & 10/59 & 0.16 / 0.17 / 0.17 & none, $p=0.98$ & no \\
\bottomrule
\end{tabular}
\end{table}

\subsection{The critic agrees with measurement when it can see what is measured}
\label{sec:res-critic}

\begin{figure}[t]
\centering
\includegraphics[width=0.95\textwidth]{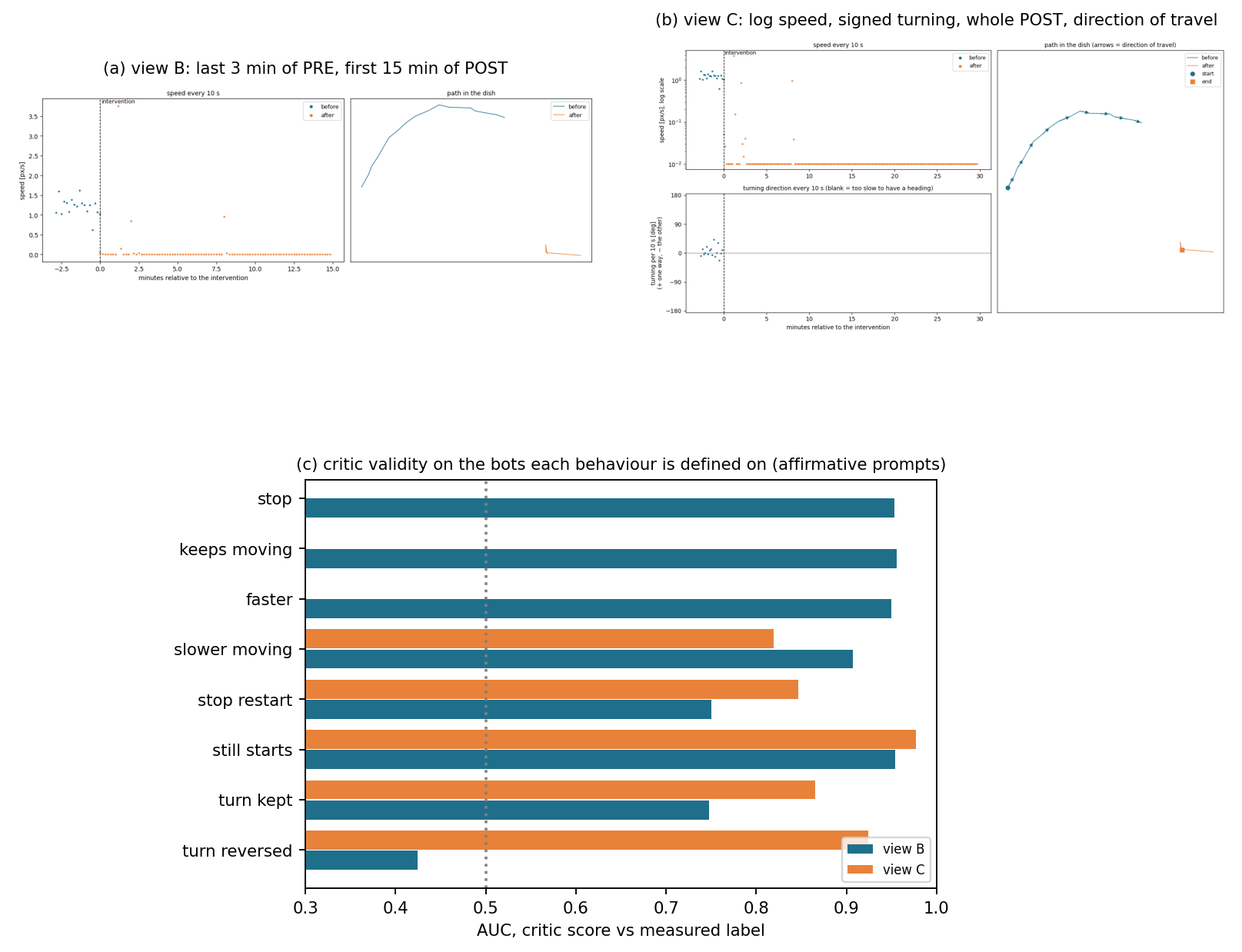}
\caption{The blind critic. The same bot as Figure~\ref{fig:overview}, as shown in (a) view B
and (b) view C. (c) AUC of the critic's score against the measured label, averaged over the
affirmative instructions of each behaviour, on the bots each behaviour is defined on
(Table~\ref{tab:behaviours}); dotted line: chance.}
\label{fig:critic}
\end{figure}

With view B, critic scores separate the measured labels of the four speed behaviours on the
121 bots moving at onset: averaged over the instructions of each class, the AUC is 0.95 for
STOP, 0.96 for KEEPS\_MOVING, 0.95 for FASTER and 0.91 for SLOWER\_MOVING, and negated
instructions score as well as affirmative ones (Figure~\ref{fig:critic}c,
Table~\ref{tab:a-critic}). In the 20-bot pilot, the scores for ``freeze the organism'' and
``never let it come to rest'' correlated at Spearman $-0.96$.

For the other behaviours, view B fails: turning reversal scores below chance (AUC 0.42), and
restarting and keeping the turning direction reach only 0.75. View B does not contain what
these labels use: it stops 15 minutes after the pulse, so a late restart is invisible, and it
shows no turning direction. With view C, which adds the whole POST, a log speed axis and
signed turning, the critic's agreement rises to 0.92 for turning reversal, 0.87 for turning
kept, 0.85 for restarting and 0.98 for starting from rest. SLOWER\_MOVING drops from 0.91 to
0.82 in view C, because its label is defined on 0--15 minutes while view C shows the whole
POST. In every case the critic agreed with the measurement once it was shown the information
the measurement uses.

Where the critic and the labels still disagreed, the disagreement pointed to the labels. Our
stop label counts any 3-minute window below the speed floor, so a bot that paused for two
minutes after the pulse and then swam fast was labelled as stopped; the critic did not call it
``frozen''. For bots that zig-zag, the turning angle per 10\,s sits near $\pm180^\circ$, where
the sign of the median is unstable; several of these bots are labelled as reversing their
turning by the metric, and view C shows the judge why such a label is fragile
(Appendix~\ref{app:label-weak}).

One limit concerns paraphrases. A judge scoring all instructions of an image in one session
gives nearly identical scores to paraphrases of the same class (within-image standard
deviation about 1 point, against 15--35 points between bots). The critic thus provides one
score per behaviour per bot; variation between phrasings is not measured by this protocol.

\subsection{Dose curves learned from the critic match those learned from measurement}
\label{sec:res-curves}

\begin{figure}[t]
\centering
\includegraphics[width=\textwidth]{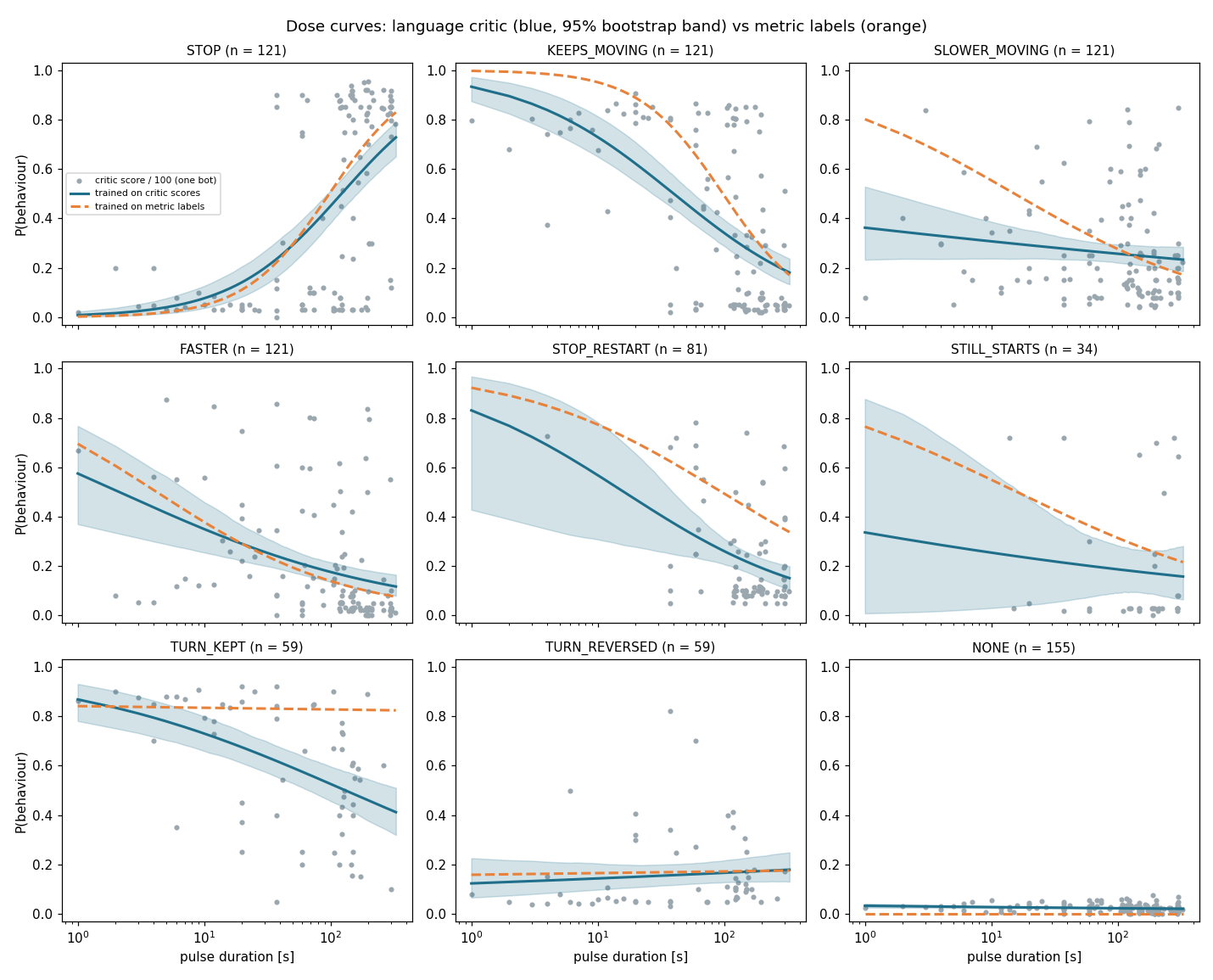}
\caption{Dose curves for all behaviours, learned from critic scores (blue, 95\% bootstrap band)
and from measured labels (orange, dashed). Grey: mean critic score of each bot. View B scores
for STOP, KEEPS\_MOVING, FASTER and SLOWER\_MOVING; view C for the others. ``NONE'' is the
out-of-scope class.}
\label{fig:curves}
\end{figure}

Fitting Equation~\ref{eq:curve} per behaviour to critic scores, with no measured label,
recovers the dose dependence measured in Section~\ref{sec:res-pulse} (Figure~\ref{fig:curves}).
For STOP the slope is positive (1.23 per standard deviation of log duration, 95\% bootstrap
interval 0.96--1.57); for KEEPS\_MOVING and FASTER it is negative ($-0.89$ and $-0.50$). On
unseen bots, these critic-trained curves predict the measured label with AUC 0.82 (STOP), 0.81
(KEEPS\_MOVING) and 0.72 (FASTER), and they beat a constant model in held-out cross-entropy.
At the planned dose, the probabilities from critic and from measurement are close: STOP at
330\,s, 0.73 (0.65--0.79) against 0.83; KEEPS\_MOVING at 1\,s, 0.93 against 1.00; FASTER at
1\,s, 0.58 against 0.70. Critic curves are slightly lower because the critic scores
borderline bots in the middle of the scale.

For the five other behaviours, critic curves from view C agree with measured curves at the
planned dose (STOP\_RESTART 0.83 against 0.92; turning kept 0.87 against 0.84; turning reversed
0.18 against 0.18), but duration does not predict them: held-out AUC 0.54 for restarting, 0.36
for turning reversal and 0.21 for starting from rest. The critic and the measurement, two
independent instruments, reach the same conclusion: these behaviours are visible, but not set
by the pulse.

\subsection{P2I generalises to unseen instructions and unseen bots}
\label{sec:res-p2i}

\begin{figure}[t]
\centering
\includegraphics[width=0.49\textwidth]{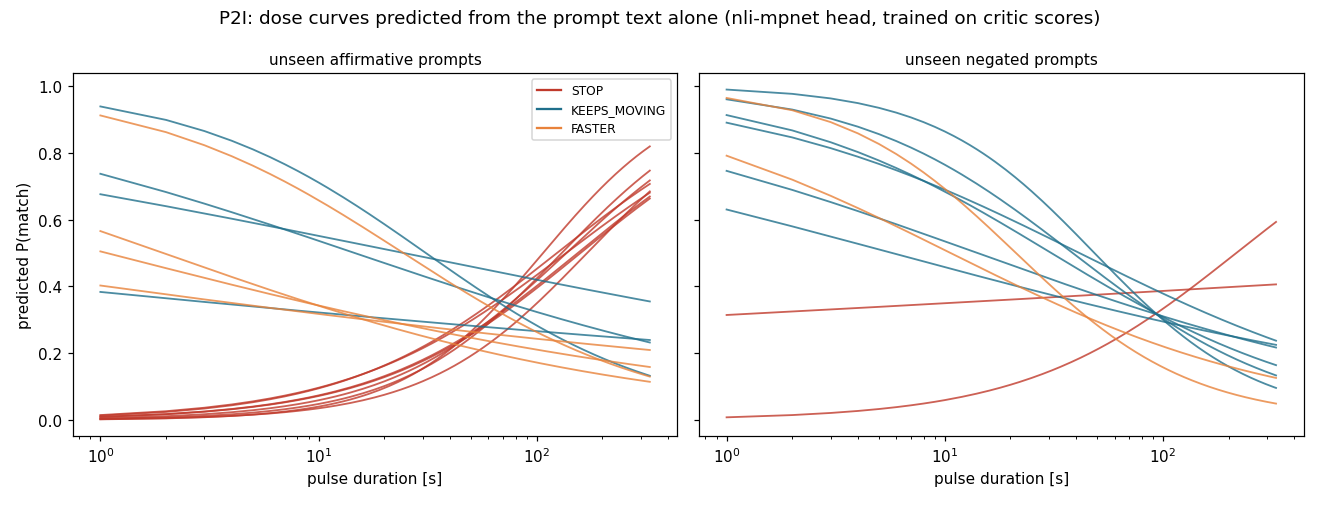}\hfill
\includegraphics[width=0.49\textwidth]{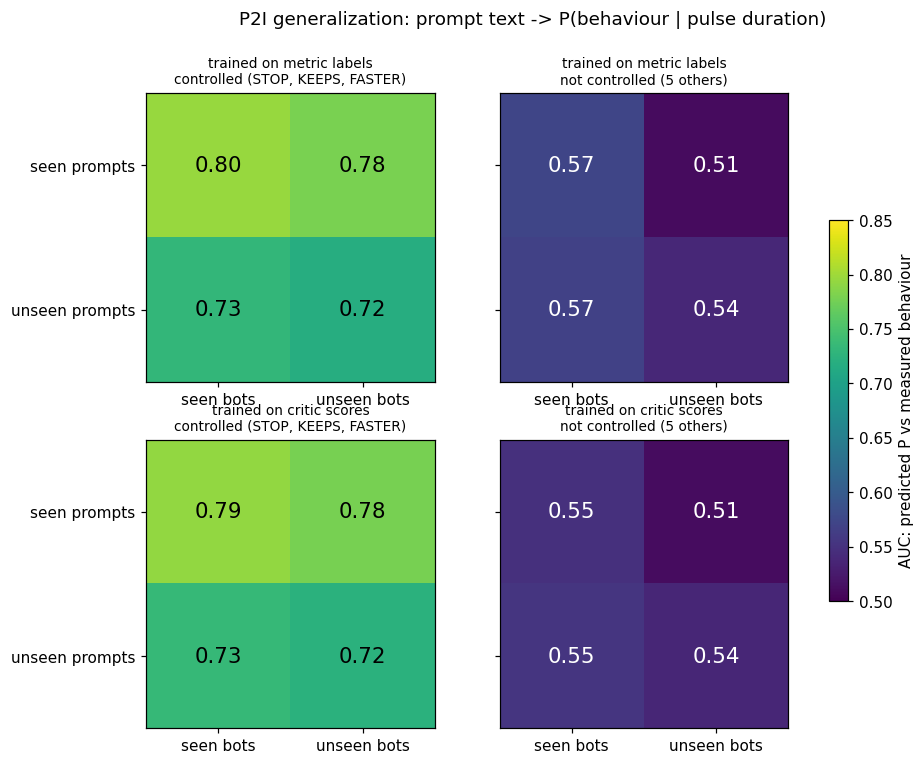}
\caption{(a) Dose curves predicted for unseen instructions of the three controlled behaviours
(linear head, critic target), affirmative and negated. (b) Generalisation over all eight
behaviours: AUC on seen or unseen instructions and seen or unseen bots, for the controlled
behaviours and for the five behaviours the pulse does not set, with measured-label and critic
targets.}
\label{fig:p2i}
\end{figure}

On the three controlled behaviours, P2I chooses the right end of the dose range for every
unseen instruction, including every negation (Table~\ref{tab:p2i}; Figure~\ref{fig:p2i}a). Its
predicted probabilities separate the measured behaviour of unseen bots with AUC 0.76 for
unseen instructions and unseen bots, against 0.79--0.80 for training instructions on training
bots. Each axis alone costs 0.01--0.03 AUC (new instructions 0.021 and 0.026, new bots 0.024
and 0.012, for measured and critic targets), so neither new wording nor new organisms dominate
the loss.

Training on critic scores and training on measured labels give the same controller: AUC
0.758 against 0.761 on unseen instructions and bots, and the right dose for all 26 unseen
instructions in every seed with either target. The critic can replace hand-made labels as the
training signal without measurable loss. A linear probe of the same embeddings, trained on
critic scores, recovers 86\% of the gap in held-out cross-entropy between a model that ignores
the instruction and one that is told the true behaviour class (Appendix~\ref{app:p2i-extra}).

\begin{table}[t]
\centering
\small
\caption{P2I on the three controlled behaviours (MLP head, nli-mpnet, five seeds; mean $\pm$
SD). AUC: predicted probability against the measured behaviour. Right dose and negations are
summed over seeds (26 unseen instructions, 10 negated, per seed).}
\label{tab:p2i}
\begin{tabular}{lcccccc}
\toprule
& \multicolumn{2}{c}{seen instructions} & \multicolumn{2}{c}{unseen instructions} & & \\
training target & seen bots & unseen bots & seen bots & unseen bots & right dose & negations \\
\midrule
measured labels & $0.802{\pm}0.001$ & $0.778{\pm}0.004$ & $0.781{\pm}0.000$ & $0.761{\pm}0.003$ & 130/130 & 50/50 \\
critic scores & $0.792{\pm}0.001$ & $0.780{\pm}0.006$ & $0.766{\pm}0.001$ & $0.758{\pm}0.004$ & 130/130 & 50/50 \\
\bottomrule
\end{tabular}
\end{table}

When all 176 instructions and eight behaviours are learned by one model
(Figure~\ref{fig:p2i}b), the three controlled behaviours keep AUC 0.72 on unseen instructions
and bots with either target, while the five uncontrolled behaviours stay near chance
(0.51--0.57). This is the expected outcome, and a useful one: an interface built on this
archive can predict which requests the pulse can satisfy and which it cannot.

\subsection{Model capacity is not the bottleneck}
\label{sec:res-ablation}

Table~\ref{tab:ablation} compares a linear head, the MLP head, and LoRA fine-tuning of the
encoder (rank 8 on all attention projections), each trained with hyperparameters fixed before
running and repeated over five seeds that change initialisation, dropout and bot folds. The MLP
is better than the linear head in every seed, by 0.013 (measured labels) and 0.008 (critic).
LoRA is no better than the MLP and, in every seed, maps the double negation ``don't let it keep
swimming'' to the wrong end of the range. With 114 training sentences, the information that
transfers to new wording is already in the frozen embedding; more trainable capacity does not
add to it. Encoders not trained for inference fail on negation: with all-MiniLM-L6-v2, the
right dose is chosen for 17--18 of 26 unseen instructions and 5 of 10 negations.

\begin{table}[t]
\centering
\small
\caption{Head ablation, unseen instructions $\times$ unseen bots, five seeds. $\Delta$: paired
difference to the linear head (95\% $t$-interval).}
\label{tab:ablation}
\begin{tabular}{llcccc}
\toprule
target & head & AUC & $\Delta$ vs linear & right dose & negations \\
\midrule
measured & linear & $0.748{\pm}0.003$ & -- & 125/130 & 45/50 \\
measured & MLP & $0.761{\pm}0.003$ & $+0.013$ [0.011, 0.014] & 130/130 & 50/50 \\
measured & LoRA & $0.752{\pm}0.004$ & $+0.004$ [0.002, 0.007] & 125/130 & 45/50 \\
critic & linear & $0.750{\pm}0.004$ & -- & 130/130 & 50/50 \\
critic & MLP & $0.758{\pm}0.004$ & $+0.008$ [0.006, 0.009] & 130/130 & 50/50 \\
critic & LoRA & $0.759{\pm}0.005$ & $+0.008$ [0.007, 0.010] & 125/130 & 45/50 \\
\bottomrule
\end{tabular}
\end{table}

\section{Discussion and limitations}
\label{sec:limits}

An archive of experiments collected for other purposes supports a small but honest language
interface to a living organism. It can stop a xenobot, leave it moving, or favour faster
motion, choose the dose for an instruction it has never seen, including negated ones, and
attach a probability calibrated on unseen bots. It can also say which requests the archive
cannot serve: turning direction, restarting and starting from rest are visible to both the
critic and the measurements, but the pulse duration does not predict them on unseen bots. A blind language critic can
replace hand-made labels, on one condition: its plots must contain the information the
behaviour is defined on. When they did not, the critic failed, and the failure was diagnosable;
when they did, the critic also exposed flaws in our own labels.

\paragraph{Limitations.}
\begin{itemize}
\item \textbf{No sham control.} The archive has no unstimulated recordings, and constructed
references were shown to be invalid. All claims are associations with dose and predictions on
unseen bots, not effects of a pulse against no pulse. Sham trials are the first requirement for
future experiments.
\item \textbf{One intervention parameter.} Duration is the only control, and the monotone dose
curves make the plan effectively ``short or long''. A planner that returns the shortest duration
reaching a target probability, $d^\star = \exp\!\big(\mu + \tau\,\mathrm{logit}(p^\star)\big)$,
and refuses when no duration in the archive reaches it, gives continuous doses and refusals
from the same model; testing it requires new experiments.
\item \textbf{No prospective test.} Planned doses were evaluated on held-out archive bots, not
applied to new bots.
\item \textbf{The critic.} Each image was scored once, by one model; repeated scoring, other
judges and human raters are needed to measure its stability. Scores for paraphrases scored in
one session are anchored to each other.
\item \textbf{Instructions.} 176 instructions written by the authors; the main uncertainty in
P2I results is which instructions were written, not training randomness. An independently
written test set would sharpen the comparison between heads.
\end{itemize}

\paragraph{Reproducibility.}
Tracks, measured outcomes, every critic prompt and response with hashes, instruction sets and
all analysis code will be released with the paper.

\section*{Statement on the use of AI}
A multimodal language model (Claude Opus 5.5, Anthropic) is a methodological component: it is
the blind critic of Section~\ref{sec:critic}, and its agreement with independent measurements is
evaluated rather than assumed. An AI assistant was also used for code and text editing. All
content was reviewed by the authors, who take responsibility for it.

\bibliographystyle{plainnat}
\bibliography{refs}

\newpage
\appendix
\section{Tracking}
\label{app:tracking}

\paragraph{Failure of a segmentation tracker.}
The segmentation tracker prompted the segmenter with the last known position, accepted a mask
by a pixel-count test and a colour check, fell back to optical flow and a grid search, and
otherwise held the last box. An audit of its own signals, without any outcome data, found
23.8\% of frames held and 6.2\% re-acquired by fallbacks whose median step was about one body
length per 10\,s (0.035 for direct detections), 39\% of which jumped out and straight back.
Rendering recordings 135 and 367 at the moment of loss showed thin-band masks
($6050\times122$ and $6195\times875$ pixels) with the bot at one end.

\paragraph{Classical tracker.} Decoding at 1/4 resolution, darkness is a Gaussian-blurred frame
($\sigma = 15$ pixels) minus the frame; candidates are connected regions more than 25 grey
levels darker. A candidate must match the bot's area within 0.25--4$\times$, its box within
2.5$\times$, have aspect ratio at most 3, peak darkness at least half the bot's, a dark centre
(rejecting hollow bubbles), and a floor at least 0.6 times as bright as the bot's surroundings
(rejecting the electrodes and rim). Over all 163 recordings it finds the bot near its last
position in 94.2\% of frames, re-acquiring it by global search 51 times.

\paragraph{Robust PCA.} Inexact augmented Lagrangian at 1/16 resolution. Hold rule: when the
bot leaves the sparse layer, its last position is kept and the raw frame is checked for a dark
object (held-dark) or not (held-empty, lost). On a recording in which the bot stops 0.4 minutes
after the pulse, all 323 held frames lie within half a body length of the classical tracker.

\begin{figure}[H]
\centering
\includegraphics[width=\textwidth]{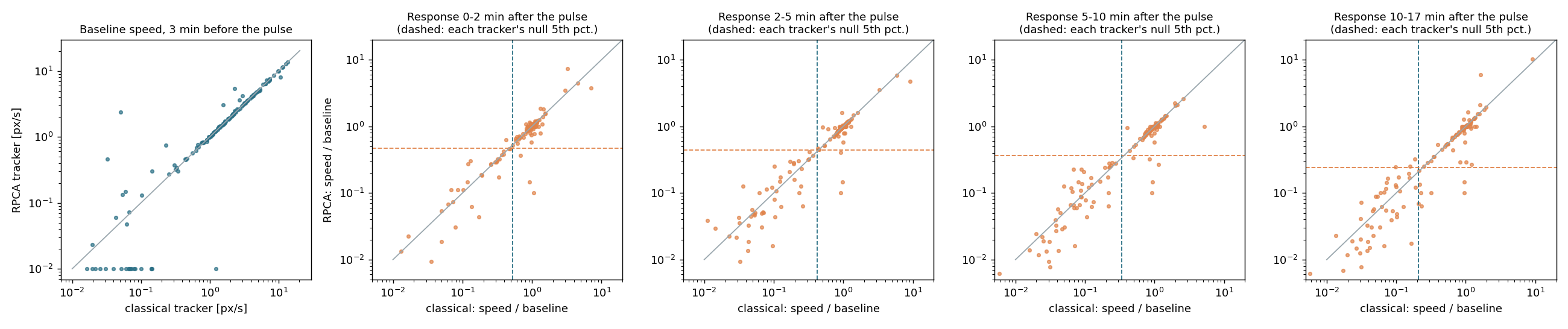}
\caption{The two trackers compared on all recordings: baseline speed (left) and response
relative to baseline in four windows after the pulse.}
\label{fig:a-trackers}
\end{figure}

\section{Baseline length}
\label{app:sensitivity}

\begin{figure}[H]
\centering
\includegraphics[width=0.6\textwidth]{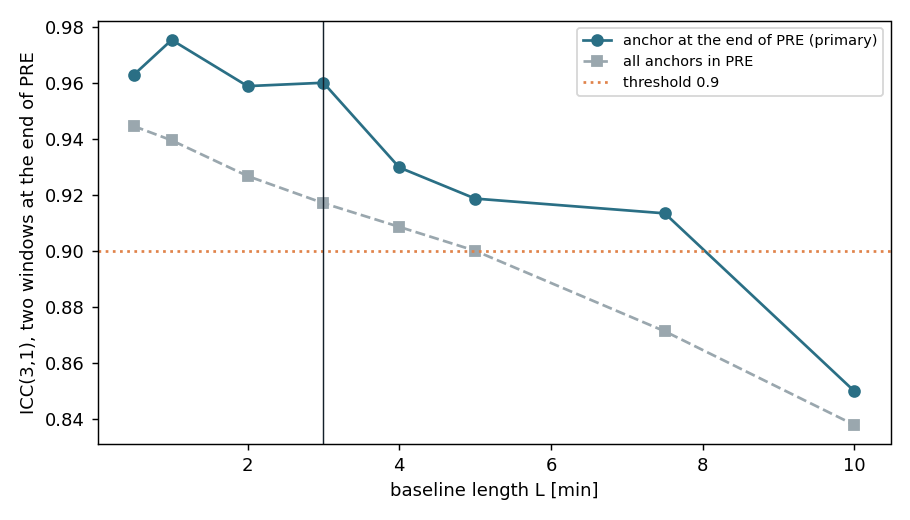}
\caption{Agreement of two adjacent PRE windows against window length, at the end of PRE
(primary) and anywhere in PRE. The rule written before the analysis chose 3 minutes.}
\label{fig:a-baseline}
\end{figure}

\begin{table}[H]
\centering
\small
\caption{Baseline length, PRE frames only, 128 bots moving before the pulse.}
\begin{tabular}{rccc}
\toprule
length (min) & ICC at end of PRE & 90\% drift & usable \\
\midrule
0.5 & 0.963 & $\times1.72$ & 98\% \\
1 & 0.975 & $\times1.50$ & 98\% \\
2 & 0.959 & $\times1.71$ & 97\% \\
\textbf{3} & \textbf{0.960} & $\times\mathbf{1.45}$ & 98\% \\
5 & 0.919 & $\times1.84$ & 98\% \\
10 & 0.850 & $\times2.98$ & 66\% \\
\bottomrule
\end{tabular}
\end{table}

A bootstrap over bots selects 3 minutes in 57\% of resamples and 1 minute in 32\%. With 1, 2
and 5-minute baselines the Cox hazard ratio per $e$-fold of duration is 2.71, 2.61 and 2.64,
and the fraction still moving 10 minutes after pulses above 220\,s is 0.08--0.09 at every
length.

\section{Constructed controls do not work for these organisms}
\label{app:placebo}

Fake onsets placed inside PRE, where nothing happens, should be flagged as significant 5\% of
the time by a calibrated test. Per-bot nulls built from each bot's own PRE history flag
21--23\% per side; parametric event studies flag 27--62\%. Bots change their behaviour over
tens of minutes, so PRE segments are not exchangeable and cannot serve as a sham.

\section{Critic templates and instruction sets}
\label{app:templates}

The view-B template, verbatim except for the instruction list and file paths:

\begin{quote}\small\ttfamily
You are an independent judge. You see one figure of a single xenobot, a small multicellular
organism that swims in a dish. The left panel shows its speed every 10 seconds; the right panel
shows its path in the dish. Blue is before an intervention, orange is after it; the dashed line
marks the intervention.\\
Step 1. Look at the figure and describe, in 2 to 4 plain sentences, what the bot does before
the intervention and what it does after it.\\
Step 2. For each numbered instruction below, judge how well the bot's behaviour AFTER the
intervention, compared with its own behaviour BEFORE it, matches what the instruction asks for.
Use only what the figure shows. Do not assume anything about the intervention itself. If an
instruction asks for something the figure cannot show, or something the bot clearly did not
do, score it low.\\
Score each instruction from 0 to 100: 0 = clearly does not match, 50 = cannot tell, 100 =
clearly matches. [\ldots]
\end{quote}

The view-C template differs only in the description of the panels (log speed, signed turning,
arrows for the direction of travel). The instruction set has 176 instructions: per behaviour,
12--15 training and 4--8 test instructions, with 2 negated test instructions per behaviour (6
for KEEPS\_MOVING), and 18 out-of-scope instructions.

\begin{table}[H]
\centering
\small
\caption{Critic validity by behaviour (mean AUC over instructions; affirmative / negated).}
\label{tab:a-critic}
\begin{tabular}{lcc}
\toprule
behaviour & view B & view C \\
\midrule
STOP & 0.953 / 0.953 & -- \\
KEEPS\_MOVING & 0.956 / 0.960 & -- \\
FASTER & 0.950 / 0.957 & -- \\
SLOWER\_MOVING & 0.907 / 0.917 & 0.819 / 0.831 \\
STOP\_RESTART & 0.750 / 0.730 & 0.847 / 0.827 \\
STILL\_STARTS & 0.954 / 0.954 & 0.977 / 0.977 \\
TURN\_KEPT & 0.748 / 0.752 & 0.865 / 0.870 \\
TURN\_REVERSED & 0.424 / 0.531 & 0.924 / 0.882 \\
\bottomrule
\end{tabular}
\end{table}

\paragraph{Pilot (20 bots).} Four instructions, each image scored by a fresh judge, view A then
view B with new codes and orders. AUCs against the measured labels were 0.92/0.92 (stop),
0.93/0.99 (keeps moving), 1.00/1.00 (faster, 2 positives), 0.99/0.94 (slower, still moving);
scores agreed across views at Spearman 0.73--0.97. One bot labelled ``stopped'' scored low on
``freeze'' in view A (35/100) because, over the whole PRE, it had already been mostly at rest;
in view B, which shows only the last 3 minutes of PRE, it scored 72.

\section{Where the labels were weak}
\label{app:label-weak}

\paragraph{Transient stops.} The stop label counts any 3-minute window below 0.1\,px/s. Recording
114 pauses for about two minutes right after the pulse and then swims as fast as before; it is
labelled ``stopped at 0 minutes''. The judge scored ``freeze'' 15--20 and described a pause
followed by fast swimming. A persistence condition would remove such cases.

\paragraph{Turning near $\pm180^\circ$.} For bots that reverse heading almost every 10\,s, the
turning angle sits near $\pm180^\circ$, where the median's sign is unstable; in recording 093,
labelled as reversing its turning direction, view C shows a zig-zag rather than a change of
turning direction.

\section{No new behaviour after the pulse}
\label{app:discovery}

Every recording was cut into 3-minute segments (sliding by 1 minute) in PRE and POST, each
described by seven trajectory features (log speed, speed spread, share of still steps,
straightness, curvature, consistency of turning, heading memory). A segment's novelty is its
distance to the fifth-nearest PRE segment of other bots; POST segments beyond the 99th
percentile of PRE novelty were 0.7\% of POST segments, below the 1\% expected by chance.

\begin{figure}[H]
\centering
\includegraphics[width=0.55\textwidth]{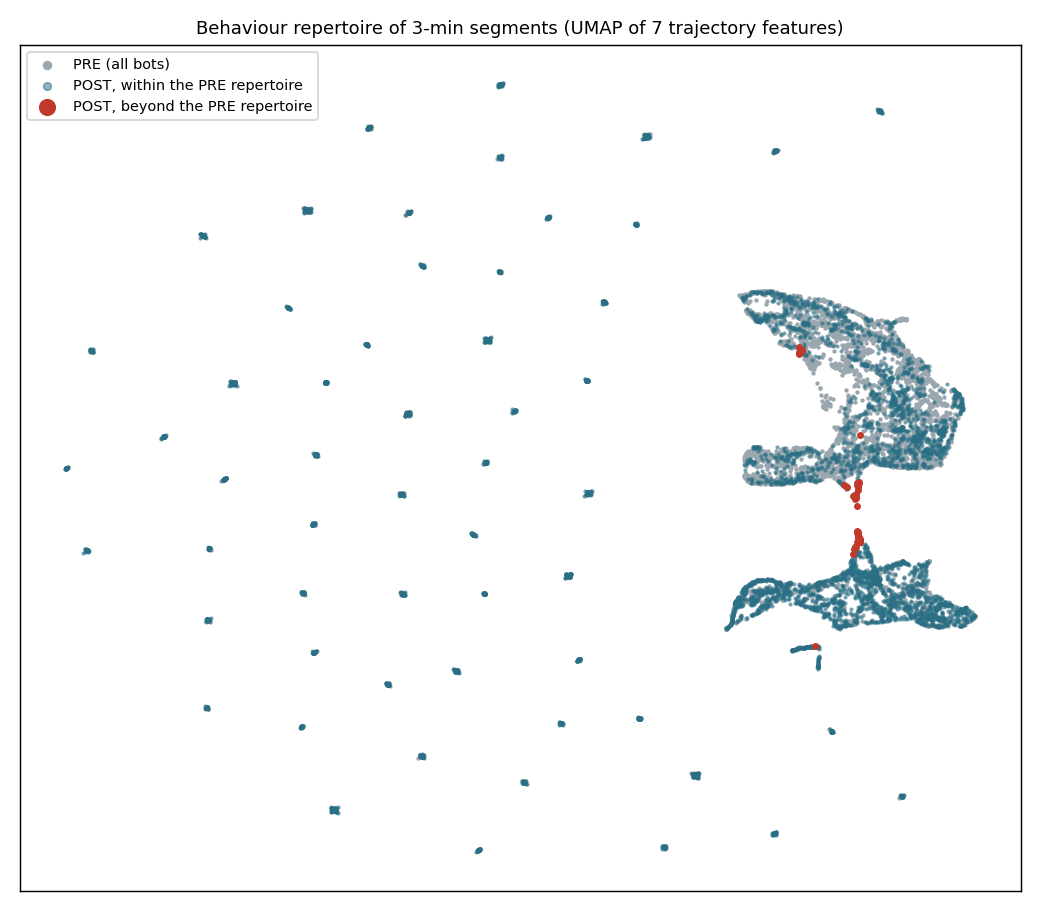}
\caption{Behaviour repertoire of 3-minute segments (UMAP of seven trajectory features).}
\end{figure}

\section{Additional P2I results}
\label{app:p2i-extra}

\paragraph{Linear probe, three controlled behaviours.} Held-out cross-entropy on unseen
instructions and bots: 0.575 (critic target) against 0.653 for a model that ignores the
instruction and 0.562 for one told the true class. With the measured target the right dose is
chosen for 25 of 26 unseen instructions (9 of 10 negations); the single error is the double
negation ``don't let it keep swimming''.

\paragraph{Why the head outputs a curve, not a duration.} Regressing log duration directly from
the embedding, with each training instruction's planned duration as target, gave a lower success
probability on test instructions (0.58 against 0.76 for routing through the behaviour's dose
curve, over nine classes) and cannot express uncertainty or refuse.

\end{document}